\documentstyle[twoside,fleqn,espcrc2]{article}

\newcommand{\be}{\begin{equation}}
\newcommand{\ee}{\end{equation}}
\newcommand{\half}{\frac{1}{2}}

\newcommand{\AmS}{{\protect\the\textfont2
  A\kern-.1667em\lower.5ex\hbox{M}\kern-.125emS}}

\title{Mass Spectrum and Bounds on the Couplings in Yukawa
       Models With Mirror-Fermions}

\author{L.~Lin, G.~M\"unster, M. Plagge
        \vskip\baselineskip
        Inst. f. Theor. Physik I,
        Universit\"at M\"unster  \\
        Wilhelm-Klemm-Str. 9, 4400 M\"unster, Germany
         \vskip\baselineskip
         I.~Montvay, H.~Wittig\thanks{Present address: Physics
         Department, The University, Southampton, SO9 5NH, UK}
         \vskip\baselineskip
        Deutsches Elektronen-Synchrotron DESY \\
        Notkestr. 85, 2000 Hamburg 52, Germany
         \vskip\baselineskip
        C. Frick, T. Trappenberg
         \vskip\baselineskip
        HLRZ, P.O. Box 1913, 5170 J\"ulich, Germany\\}

\begin{document}

\begin{abstract}
The $\rm SU(2)_L\otimes SU(2)_R$ symmetric
 Yukawa model with mirror-fermions in the limit where
the mirror-fermion is decoupled is studied both analytically and
numerically. The bare scalar self-coupling $\lambda$ is fixed at zero
and infinity. The phase structure is explored and the relevant phase
transition is found to be consistent with a second order one. The
fermionic mass spectrum close to that transition is discussed and a
first non-perturbative estimate of the influence of fermions on the
upper and lower bounds on the renormalized scalar self-coupling is
given. Numerical results are confronted with perturbative predictions.
\end{abstract}

\maketitle

\section{INTRODUCTION}

This contribution represents the talks ``Phase Structure of a Chiral
SU(2) Yukawa Model", ``Mass Spectrum and Bounds on the Couplings in
Yukawa Models with Mirror-Fermions" delivered at the international
conference on lattice field theory ``Lattice 92", Amsterdam,
15--19 September 1992, by Lee Lin and Hartmut Wittig, respectively.

The minimal Standard Model (SM) has been very successful. So far all the
experimental data are in good agreement with its perturbative
predictions, but there are still too many free parameters (18, if all
neutrinos are massless) in the SM, hence physicists believe that it is
at most a good low energy effective theory, and new physics which cannot
be described by the SM will have noticeable effects at a higher
energy scale. Here, we concentrate on two free parameters in the SM: the
top quark mass and the mass of the Higgs
particle which is the remnant scalar particle after spontaneous symmetry
breaking. We would like to investigate whether these two
parameters are completely free, or whether there
are some limitations on their values. Hopefully, one can
also get some hints about the new physics by studying this
Higgs-fermion sector in the SM.

Since the presently quoted upper bound on the top quark mass of 200 GeV
is a one-loop perturbative result, nonperturbative (i.e. lattice)
studies of this issue can be very helpful.
In order to deal with the problem of fermion doublers in lattice
formulations, we take the mirror-fermion approach~\cite{IM}.

 Our calculations were done in the SU(2) version of the Yukawa model
with explicit mirror pairs of fermion doublet fields. The lattice action
is a sum of the O(4)
 ($\cong \rm SU(2)_L \otimes SU(2)_R$) symmetric pure scalar part
 $S_\varphi$ and the fermionic part $S_\Psi$:
\be \label{eq01}
S = S_\varphi + S_\Psi \ .
\ee
 $\varphi_x$ is the $2 \otimes 2$ matrix scalar field, and
 $\Psi_x \equiv (\psi_x, \chi_x)$ stands for the mirror pair of fermion
 doublet fields where $\psi$ is the fermion doublet and $\chi$ the
 mirror-fermion doublet.
 In the usual normalization conventions for numerical simulations we
 have
$$
S_\varphi = \sum_x \bigg\{ \half {\rm Tr\,}(\varphi_x^+\varphi_x) +
\lambda \left[ \half{\rm Tr\,}(\varphi_x^+\varphi_x) - 1\right]^2
$$
\be
\,\,\,\,  - \kappa\sum_{\mu=\pm1}^{\pm4}
{\rm Tr\,}(\varphi^+_{x+\hat{\mu}}\varphi_x)
\bigg\} \ ,
\ee
$$
S_\Psi = \sum_x \bigg\{ \mu_{\psi\chi} \left[
(\overline{\chi}_x\psi_x) + (\overline{\psi}_x\chi_x) \right]
$$
$$
- K \sum_{\mu=\pm 1}^{\pm 4} \Big[
(\overline{\psi}_{x+\hat{\mu}} \gamma_\mu \psi_x) +
(\overline{\chi}_{x+\hat{\mu}} \gamma_\mu \chi_x)
$$
$$
+ r \left[ (\overline{\chi}_{x+\hat{\mu}}\psi_x)
- (\overline{\chi}_x\psi_x)
+ (\overline{\psi}_{x+\hat{\mu}}\chi_x)
- (\overline{\psi}_x\chi_x)  \right] \Big]
$$
$$
+ G_\psi \left[ (\overline{\psi}_{Rx}\varphi^+_x\psi_{Lx}) +
(\overline{\psi}_{Lx}\varphi_x\psi_{Rx}) \right]
$$
\be
+ G_\chi \left[ (\overline{\chi}_{Rx}\varphi_x\chi_{Lx}) +
(\overline{\chi}_{Lx}\varphi^+_x\chi_{Rx}) \right] \bigg\} \,\, .
\ee
 Here $K$ is the fermion hopping parameter, $r$ the Wilson-parameter,
 which will always be fixed to $r=1$, the
 indices $L,R$ denote the chiral components of fermion
 fields, and Tr acts on the SU(2) indices only.
 In this normalization the fermion mirror-fermion mixing mass is
 $\mu_{\psi\chi}=1-8rK$. The lattice spacing $a$ is set to unity.

At $G_\chi=0$, the action has the Golterman-Petcher
shift-symmetry such that all  higher
vertex functions containing the $\chi$-field vanish
identically~\cite{GOLPET,LW}. In this limit, the $\chi$-$\chi$
and $\chi$-$\psi$ components of the two-point fermion vertex function
(the inverse fermion propagator) $\tilde\Gamma_\Psi(p)$ are equal to the
corresponding components of the free inverse propagator~\cite{Fetal}.
By setting $K=1/8$, one can easily show that in the broken phase,
there is no mixing between the fermion and mirror-fermion and
the mirror-fermion completely decouples like a right-handed neutrino.
This $G_\chi=0$, $K=1/8$ combination is sometimes
called the mirror-fermion decoupling limit. This leaves us only one
mass parameter $\kappa$ to tune when we try to approach the continuum
limit and hence saves a lot of CPU time. Numerical simulations can
also be guided by analytic results derived in this limit such that
fluctuations can hopefully be reduced. The Hybrid Monte Carlo
algorithm~\cite{HMC} is used in our simulations.
We therefore need to double the flavour of the fermion spectrum to
guarantee a positive definite fermion matrix determinant.

One of the main goals in the analysis of the model is the determination
of the range of renormalized quartic couplings~$g_R$ that can be realized
for a certain value of the cut-off as a function of the renormalized
Yukawa coupling~$G_{R\psi}$. This so-called ``Allowed Region" is
bounded by the (upper) triviality bound on $g_R$ (saturated at
$\lambda=\infty$) and the (lower) vacuum stability bound, which is
defined at $\lambda=0$. Hence for the bare quartic coupling~$\lambda$
the values $\lambda=\infty, 0$ are crucial in the numerical simulations.

Reflection positivity, which is required for any Euclidean quantum field
theory respecting unitarity, can be proven to hold in a wide range of the
parameter space of our model~\cite{LMMW}. In particular, reflexion
positivity only holds for $\kappa\geq0$. Although the criterion of
reflexion positivity is only a sufficient but not a necessary condition
for the reconstruction of the theory in Minkowski space, we
would like to stay where unitarity is guaranteed in order to be on the
safe side. So we always study the ``Allowed Region" for the renormalized
couplings in the subspace where $\kappa$ is non-negative.

The $\beta$-functions have been calculated up to one loop on the
lattice and two loops in the continuum~\cite{LW}. There are basically
two possibilities for the cut-off dependence of the ``Allowed Region":
either this region shrinks to the origin as the cut-off grows or it
expands and eventually fills the whole space of renormalized couplings
at infinite cut-off. The first possibility is due to the fact that there
is only one fixed point at zero couplings and that fixed point is
infrared stable. One-loop $\beta$-functions are behaving qualitatively
like this. In the second scenario, there is an ultraviolet stable fixed
point at nonzero couplings which changes the dependence of the
``Allowed Region" on the cut-off qualitatively. Assuming that the
$\beta$-functions are behaving qualitatively like at one-loop level,
then our model is trivial in the continuum limit, and the upper bound on
the renormalized scalar self-coupling $g_R$ will be given by setting
$\lambda=\infty$ while the lower bound comes from the
$\lambda\rightarrow 0$ limit~\cite{LMMW}.

\section{PHASE STRUCTURE}

In order to know where and how the cut-off can be removed, we need to
explore the phase structure. If there is a first order phase transition
somewhere in the bare parameter space, we would like to know how it
is going to affect the ``Allowed Region".

We basically use the
magnetization $\langle|\phi|\rangle$ and staggered
magnetization $\langle|\hat\phi|\rangle$ as the order parameters to
distinguish different phases where $\langle\,\,\rangle$ is the
statistical average, and
$$
|\phi|\equiv\sqrt{\phi_a^2}\,\, ,\,\,
|\hat\phi|\equiv\sqrt{\hat\phi_a^2}\,\, ,
$$
$$
\phi_a\equiv {1\over{L^3\cdot T}}\sum_x\,\phi_a(x)\,\, ,\,\,
$$
$$
\hat\phi_a\equiv {1\over{L^3\cdot T}}\sum_x\,(-1)^x\,\phi_a(x)\,\, ,
\,\,a=1,\ldots,4
$$
where $L$ and $T$ are lattice sizes along the spatial and time
directions. The symmetry broken (FM) phase has nonzero
$\langle|\phi|\rangle$ and vanishing $\langle|\hat\phi|\rangle$, the
symmetric (PM) phase has zero $\langle|\phi|\rangle$ and
$\langle|\hat\phi|\rangle$ while
$\langle|\phi|\rangle=0$, $\langle|\hat\phi|\rangle\ne 0$ in the
anti-ferromagnetic (AFM) phase.
In the PM phase, the mirror-fermion and fermion are degenerate and
hence it is not physical. The physically relevant phase is the FM phase
where the splitting between fermion and mirror-fermion masses is possible
due to spontaneous symmetry breaking and by tuning the bare couplings
appropriately. The AFM phase is
like the FM phase except that the staggered scalar field $\hat\phi$ is
now playing the role of $\phi$.

The phase structure can be investigated analytically using various
expansions in several limits~\cite{LAT90}. We find that
at zero or infinite Yukawa couplings, or at $K= \infty$, the system
goes to a purely scalar four-component $\phi^4$ theory plus free
fermions. Therefore, the system exists in the FM, PM and AFM phases
with the transitions between them being second order and Gaussian.
At $K=0$, which is the limit where the bare fermion mass is infinite,
a first order phase transition line was found in the U(1) version at
$G_\psi\cdot G_\chi >0$ and finite $\lambda$~\cite{LJM}. This
first order transition is due to the singularity of a log-term in the
effective action. It becomes weaker and weaker and eventually vanishes
at $\lambda=\infty$. In the SU(2) version, everything is qualitatively
the same.
Therefore, this first order transition is also present when $G_\psi\cdot
G_\chi >0$, $\lambda=$~finite. We did not further investigate this
issue. At $\lambda=\infty$, when Yukawa couplings are weak or strong,
one can do expansions up to the next-leading order in $|G|$ or
$1/|G|$ plus the small-K expansion and find out that $\kappa_c$, the
value where the transition between the FM and PM phases happens, will
go down as  $|G|$ or $1/|G|$ increase. Furthermore the transition
remains a second order one on which the renormalized scalar
mass vanishes.

We also need to look for the fermionic critical plane on which the
renormalized fermionic mixing mass $\mu_R$
vanishes. When $G_\chi=0$ at any $\lambda$, this happens at
$K=1/8$ and any $G_\psi$ value. When both $G$'s are nonzero, we can use
one-loop bare perturbation theory to estimate its position at weak $K$
and $G$'s.

The correct continuum limit should be taken by approaching the critical
line where both scalar and fermion masses are zero from within the FM
phase while at the same time the mixing mass is fixed at zero and all
mass ratios are kept constant.

The phase structure at intermediate values of the Yukawa couplings must
be explored numerically. That was always done on the $4^3\cdot 8$
lattice. The strategy is to fix $\lambda$ at infinity or zero (actually
$10^{-6}$), which are the relevant values to studying upper and lower
limits on $g_R$ respectively. For the remaining parameters,
since we chose to decouple the mirror-fermion like a right-handed
neutrino in the SU(2) model, we always set $G_\chi=0, K=1/8$,
and then studied the phase diagram in the ($\kappa,G_\psi$) plane.

At $\lambda=\infty$, the phase structure was explored
numerically up to $G_\psi$ around 2.0. The result was
shown in Fig.1 in ref.~\cite{Fetal}. Everything appeared to be
qualitatively the same as the phase structure of the U(1) model at
infinite $\lambda$. At $G_\psi\ge 1.5$, $\kappa=-0.5$, a new phase with
nonvanishing magnetization and staggered magnetization was found. We
call it the ferri-magnetic (FI) phase. The phase transition between the
FM and PM phases was found to be consistent with a second order one
because of the smooth behaviour of $\langle|\phi|\rangle$ across the
transition. This was also true for other transitions in the phase
diagram. Based on our analytic analyses, we know that as we move
to larger and larger $G_\psi$ values, eventually the FI phase will
disappear and the system again exists only in the FM, PM and AFM phases.
We therefore did not spend CPU time to investigate further. (This has
been numerically confirmed in our U(1) model~\cite{LMW}.)

The phase structure at very small $\lambda$ and large negative values
of $\kappa$ was actually not thoroughly explored yet. In the U(1) model
we only made sure that the FM-PM transition was consistent with a second
order one and then went on to study the lower bound on $g_R$~\cite{LMMW}.
We now spent more computer time to look into the phase structure of the
SU(2) model at $\lambda=10^{-6}$. The transitions from the FM to PM and
from the PM to AFM phases were found to be consistent with a second order
one at  weak $G_\psi$. Again the magnetization behaves smoothly across
the PM to FM transition. Plotting histograms of the magnetization we
found that there is no evidence at all for a two-state signal.
The two transition lines bend down as $G_\psi$ increases, and come quite
close to each other at $G_\psi=0.75$, $\kappa=-0.17$. At an even more
negative $\kappa$-value: $-0.19$ and at $G_\psi=0.6, 0.8$, we observed
that the system seemed to ``tunnel" from the FI phase to the AFM phase.
At this stage, we cannot decide whether it is a real tunneling or
whether the system has not equilibrated yet.
Even if we find a tunneling event (or a hysteresis loop), it still might
be due to a very long auto-correlation length in the vicinity of a
critical point and therefore is still not the final word. We think it
is better to measure the shape of the effective potential to see if a
double-well structure developes. If there is a first order phase
transition, we suspect that it might be the continuation of the
first order transition we discovered at $K=0$. On the other hand,
a leading order large-$N$ calculation of the effective potential of
our SU(2) model shows that the usual PM, FM, AFM and FI phases exist
at small~$\lambda$, and that all transitions are second order due to
the absence of a quartic term in the effective potential in leading
order. This issue will be further investigated in the future.

Most importantly, the physically relevant FM-PM phase transition at
$\lambda=10^{-6}$ is consistent with a second order one, and we can
define the continuum limit by approaching it from the FM phase.
Therefore, studies of the lower limit on $g_R$ should not be affected
by a possibly existing first order transition.

\section{MASSES AND COUPLINGS}

We now describe the Monte Carlo simulations of the model in the
broken~(FM) phase. Besides the massive component $\sigma_x$ of the
scalar field $\varphi_x$ three massless Goldstone bosons $\pi_{jx},\,
j=1,2,3$ appear which cause the strong finite-size effects encountered
in the simulations.

As noted before, the model was simulated using the Hybrid Monte Carlo
algorithm. For the fermions, periodic spatial boundary conditions
but antiperiodic boundary conditions in the time direction were
chosen. This fixes the minimum lattice momentum for fermionic
quantities at $p_{\rm min}=(0,0,0,\pi/T)$ where~$T$ is the time extent
of the lattice.
Exploiting the shift symmetry, the decoupling of the mirror-fermion
was ensured by setting $G_\chi=0$. In order to exclude mixing between
$\psi$- and $\chi$-states we chose a slightly different value than
$K=1/8$ which was originally suggested. Namely, by setting
$\mu_{p_{\rm min}}=0$, where $\mu_p=(\mu_{\psi\chi}+Kr\hat{p}^2)/2K$,
the fermionic hopping parameter $K$ turns out slightly greater than 1/8,
but will eventually approach this value once the time extent of
the lattice goes to infinity. This particular choice corresponds to
exact decoupling in the continuum limit whilst ensuring a smooth
behaviour of the propagator on a finite lattice near $K=1/8$
(see also ref.~\cite{Fetal}).

With the parameters $G_\chi$ and $K$ fixed by the decoupling condition
our further strategy was as follows. For $\lambda=\infty$, $G_\psi$ was
varied from 0.3, 0.6 to 1.0. At each value of $G_\psi$ the scalar hopping
parameter $\kappa$ was tuned in order to achieve $m_{R\sigma}\le1$.
Choosing lattice sizes of $L^3\cdot T$ of $4^3\cdot8$, $6^3\cdot12$ and
$8^3\cdot16$ it was hoped that on the largest lattice $m_{R\sigma}$ could
be brought down to about $m_{R\sigma}\simeq0.5$ in lattice units.

For $\lambda=10^{-6}$, $G_\psi$ was fixed at 0.3 and again $\kappa$ was
tuned. A second set of data, however, was generated by setting $\kappa=0$
and tuning $G_\psi$ in order to determine the maximum value of the
renormalized Yukawa coupling $G_{R\psi}$ for non-negative values of
$\kappa$, i.e. in the region where reflexion positivity  can still be
proven.

The results for the scalar mass show that $m_{R\sigma}$ is decreasing
as one approaches the critical line from above and then starts rising
again. This rise close to the phase transition is explained as a strong
finite-size effect. The minimum of $m_{R\sigma}$ increases with
increasing $G_\psi$~\cite{Fetal}. For $\lambda=\infty$, $G_\psi=0.6$,
it is necessary to have lattices as large as $8^3\cdot16$ in order to
obtain scalar masses smaller than~1 in lattice units.

At $\lambda=10^{-6}$ the tuning of $m_{R\sigma}$ is easier in the sense
that one can achieve smaller values than at $\lambda=\infty$, however
the rise of the minimum of $m_{R\sigma}$ as $G_\psi$ is increased,
remains. All that illustrates that there are strong finite-size effects
which get stronger as the bare couplings $\lambda$ and/or $G_\psi$ are
increased. Nevertheless, our experience has shown that for $\kappa\ge0$
a lattice size of $8^3\cdot16$ is sufficient to have control on the finite
volume effects.

For the fermion masses one can make the following statements: firstly,
setting $G_\chi=0$ ensured indeed that the mass of the mirror-fermion
$\mu_{R\chi}$ was always zero within errors. Secondly, the condition
$\mu_{p_{\rm min}}=0$ guaranteed that the fermionic mixing mass $\mu_R$
was zero as well. Fig.~1 shows the fermion mass $\mu_{R\psi}$
(full squares) and the lowest doubler states $\mu_{R\psi}^d$ (open
squares) and $\mu_{R\chi}^d$ (open circles) as a function of $\kappa$.
It is seen that the lowest fermion doublers have masses of 1.5 in
lattice units. Furthermore, it is evident from the figure that only at
the lowest value of $\kappa$ there is a clear separation of $\mu_{R\psi}$
and the doubler states. Since this $\kappa$-value coincided with the
minimum of the scalar mass ($m_{R\sigma}=0.75(3)$ at $\kappa=0.09$,
$\lambda=10^{-6}$, $G_\psi=0.3$ on $6^3\cdot12$), it is precisely this
data point which was subsequently included in the plot of the Allowed
Region. To summarize, we note that in general $\mu_{R\psi}$ decreases
as $\kappa$ and/or $G_\psi$ is decreased, and that the lowest doubler
state of the mirror-fermion has a mass of about 1.5, irrespective of
the actual choice of the tunable bare parameters.
%

Using the relation $\mu_{R\psi}=G_{R\psi}v_R$, where $v_R$ is the VEV
of the scalar field, it is instructive to plot $G_{R\psi}$ versus
$G_\psi$ in order to check whether the linear rise of $G_{R\psi}$
reported for the symmetric phase~\cite{LW} prevails. This is done
in Fig.~2. It is seen that there is still a linear behaviour
which, however, is much flatter as in the symmetric phase. The data
obtained at $\lambda=10^{-6}$ are slightly larger than those at
$\lambda=\infty$. The maximum value of $G_{R\psi}$ for
$\kappa\ge0$ was found at $\kappa=0$, $G_\psi=0.63$ where
$G_{R\psi}^{\rm max}=3.5\pm0,4$. This has to be confronted with the
tree unitarity bound which yields $G_{R\psi}\simeq2.5$ for
$N_f=2$. Hence the maximum value for $G_{R\psi}$ is not very much
above this bound, suggesting that for $\kappa\ge0$ the coupling
$G_{R\psi}$ cannot grow indefinitely large.
\begin{figure}[p]              \label{DOUBLB}
\vspace{7.3cm}
\caption{The fermion mass $\mu_{R\psi}$ and the lowest doubler states
plotted versus $\kappa$ at $\lambda=10^{-6}$, $G_\psi=0.3$ on
$6^3\dot12$.}
\end{figure}
\begin{figure}[p]              \label{GLINSB}
\vspace{7.3cm}
\caption{The renormalized Yukawa coupling $G_{R\psi}$ versus $G_\psi$
for $\lambda=\infty$ (open circles) and $\lambda=0$ (full squares) on
$6^3\cdot12$.}
\end{figure}
\begin{figure}[p]             \label{APAPALL}
\vspace{7.3cm}
\caption{The Allowed Region in the $(g_R,\,G_{R\psi}^2)$-plane together
with the integration of the $\beta$-functions for scalar mass
$m_{R\sigma}=1$ (solid line) and $m_{R\sigma}=0.75$ (dotted line).}
\end{figure}
\begin{figure}[p]             \label{RATIO}
\vspace{7.3cm}
\caption{The mass ratio $\mu_{R\psi}/m_{R\sigma}$ versus $G_{R\psi}$ in
comparison with 1-loop perturbative estimates for $m_{R\sigma}=0.75$
(dotted curve), $m_{R\sigma}=1$ (full curve) and $m_{R\sigma}=1.25$
(dashed curve).}
\end{figure}

Fig.~3 shows our data for the Allowed Region, together
with the perturbative expectation coming from the numerical integration
of the one-loop $\beta$-functions for cut-off's corresponding to
$m_{R\sigma}=0.75$ (dotted curve) and $m_{R\sigma}=1$ (solid curve),
respectively. All data were obtained on lattices of size $6^3\cdot12$
and $8^3\cdot16$. The point with the largest couplings was on
$6^3\cdot12$ at $\lambda=10^{-6}$, $G_\psi=0.63$, $\kappa=0$.
Despite the large error bars the data are in remarkable
agreement with the perturbative estimates even at values of $G_{R\psi}$
around or even above the tree unitarity limit. Therefore, the data
support the view that the perturbative behaviour of the bounds on the
scalar coupling observed in pure $\phi^4$ theory persists when fermions
are included.
%

Finally, Fig.~4 shows the mass ratio $\mu_{R\psi}/m_{R\sigma}$
as a function of $G_{R\psi}$, together with the perturbative prediction
from the integration of the one-loop $\beta$-functions for the upper and
lower bound, respectively. It is seen that for larger values of
$G_{R\psi}$ (where there is only a weak cut-off dependence) the data
are clustered around 0.7, which is close to the fixed point for the
corresponding ratio of couplings,
$\sqrt{3}\,G_{R\psi}/\sqrt{g_R} = 0.759\ldots$ that is reached as the
cut-off scale becomes infinite. The fixed point is indicated by the
horizontal line in the figure. For smaller $G_{R\psi}$ the data
obtained for $\lambda=\infty$ deviate from the fixed point in a fashion
that is well described by the expected cut-off dependence of the upper
bound for finite cut-off.
%
%

\section{CONCLUSIONS}

According to our results, the Higgs-Yukawa sector of the Standard Model
can indeed be studied non-perturbatively following the mirror-fermion
approach. The phase diagram exhibits a rich structure with the familiar
PM, FM, AFM and FI phases encountered in other lattice Yukawa models as
well. For both $\lambda=\infty$ and $\lambda\simeq0$ the physically
relevant phase transition from PM to FM is second order and hence
permits taking the continuum limit along this transition line. Exploiting
the Golterman-Petcher shift symmetry, numerical simulations in the
decoupling limit in the broken phase are feasible and greatly facilitated
since only $\kappa$ remains to be tuned once $G_\psi$ is fixed. The
resulting fermionic mass spectrum consists of a fermion of mass
$\mu_{R\psi}<1$, whereas $\mu_{R\chi}$ and the mixing mass $\mu_R$ vanish
identically. The doublers for both fermion and mirror-fermion receive
masses of at least 1.5 in lattice units.

The most severe limitations arise from finite-size effects on the scalar
mass $m_{R\sigma}$ in the broken phase. However, for an exploratory
study such as this a lattice size of $8^3\cdot16$ appears sufficiently
large in order to control finite-size effects.

Restricting the physical analysis to data points with $\kappa\ge0$
(i.e. data which always will permit a particle interpretation in
Minkowski space-time), it is evident that the results for the couplings
$g_R$ and $G_{R\psi}$ agree with perturbative estimates. In particular
the data for the Allowed Region suggest that the perturbative behaviour
of the bounds on $g_R$ encountered in pure $\phi^4$ theory prevails once
the effects of heavy fermions are taken into account.

During the preparation of this paper, we have accumulated about 3500
trajectories for a run at $\lambda=10^{-6}$, $G_\psi=0.25$,
$G_\chi=0.0$ and $\kappa=0.099$. It shows that with
$m_{R\sigma}=0.624(56)$, $\mu_{R\psi}=0.413(12)$,
we obtain $g_R=16\pm5$, $G_{R\psi}=1.56(7)$ and $G_{R\psi}^2=2.43(22)$.
Readers can see that the central value is slightly above
the dotted one-loop curve we show in Fig.~3. Since the scalar mass
in that data point turns out to be higher than $m_{R\sigma}=0.75$
for the dotted curve, our findings still support the perturbative
scenario for the bounds on $g_R$. Data with better statistics will be
published in a forthcoming paper.


\begin{thebibliography}{9}
\bibitem{IM} I. Montvay, Phys. Lett. \underline{199B} (1987) 89.
\bibitem{GOLPET} M.F.L. Golterman and D. N. Petcher, Phys. Lett.
\underline{225B} (1989) 159
\bibitem{LW} L. Lin and H. Wittig, Z. Phys. \underline{C54} (1992) 331.
\bibitem{Fetal} C. Frick , L. Lin, I.~Montvay, G.~M\"unster,
M.~Plagge, T.~Trappenberg and H.~Wittig, DESY Preprint 92-111.
\bibitem{HMC}
S. Duane, A. D. Kennedy, B. J. Pendleton, D. Roweth,
Phys. Lett. \underline{195B} (1987) 216.
\bibitem{LMMW} L. Lin, I. Montvay, G. M\"unster and H. Wittig,
               Nucl. Phys. \underline{B355} (1991) 511.
\bibitem{LAT90} L. Lin, I. Montvay, G. M\"unster and H. Wittig,
               Nucl. Phys. B (Proc. Suppl.) \underline{20} (1991) 601.
\bibitem{LJM} L. Lin, J. P. Ma and I. Montvay,
               Z. Phys. \underline{C48} (1990) 222.
\bibitem{LMW} L. Lin, I. Montvay and H. Wittig,
               Phys. Lett. \underline{264B} (1991) 407.
\end{thebibliography}
\end{document}